\documentclass[aps,pra,onecolumn,superscriptaddress]{revtex4}

\usepackage{epstopdf}
\usepackage{color}
\usepackage{url}
\usepackage{hyperref}
\usepackage[T1]{fontenc}
\usepackage[latin1]{inputenc}
\usepackage{float}

\usepackage{physics}
\usepackage{amssymb} 
\usepackage{graphicx} 

\usepackage[english]{babel} 
\usepackage{blindtext} 

\newcommand{\cann}[2]{c_#1(#2)} 
\newcommand{\Cdag}[3]{C_{#3}^{#1#2}} 
\newcommand{\tCdag}[3]{\tilde{C}_{#3}^{#1#2}} 
\newcommand{\adag}[2]{a^{\dagger#1}_#2} 
\newcommand{\bdag}[2]{b^{\dagger#1}_#2} 
\newcommand{\cdag}[2]{c^{\dagger#1}_#2} 
\renewcommand{\ddag}[2]{d^{\dagger#1}_#2} 
\newcommand{\Adag}[3]{A_{#3}^{#1#2}} 
\newcommand{\Bdag}[3]{B_{#3}^{#1#2}} 
\newcommand{\tBdag}[3]{\tilde{B}_{#3}^{#1#2}} 

\newcommand{\Ddag}[3]{D_{#3}^{#1#2}} 
\newcommand{\tDdag}[3]{\tilde{D}_{#3}^{#1#2}} 

\begin{document}

\title{Effects of cavity birefringence in polarisation-encoded quantum networks}

\author{E. Kassa}
\thanks{These two authors contributed equally }
\affiliation{University of Oxford, Clarendon Laboratory, Parks Road, Oxford, OX1 3PU, United Kingdom}
\affiliation{Experimental Quantum Information Physics Unit, Okinawa Institute of Science and Technology, Onna, Okinawa, 904-0495, Japan}
\author{W. J. Hughes}
\thanks{These two authors contributed equally }
\author{S. Gao}
\author{J. F. Goodwin}
\email{joseph.goodwin@physics.ox.ac.uk}
\affiliation{University of Oxford, Clarendon Laboratory, Parks Road, Oxford, OX1 3PU, United Kingdom}

	\begin{abstract}
	The generation of entanglement between distant atoms via single photons is the basis for networked quantum computing, a promising route to large-scale trapped-ion and trapped-atom processors. Locating the emitter within an optical cavity provides an efficient matter-light interface, but mirror-induced birefringence within the cavity introduces time-dependence to the polarisation of the photons produced. We show that such `polarisation oscillation' effects can lead to severe loss of fidelity in the context of two-photon, polarisation encoded measurement-based remote entanglement schemes. It is always preferable to suppress these errors at source by minimising mirror ellipticity, but we propose two remedies for systems where this cannot be achieved.
	We conclude that even modest cavity birefringence can be detrimental to remote entanglement performance, to an extent that may limit the suitability of polarisation-encoded schemes for large-scale quantum networks.

	\end{abstract}
	\maketitle

\section{Introduction}

The interference of single photons has formed the basis of many applications in quantum optics, from Young's double slit experiment to ascertain the wave-particle duality \cite{Young:04}, to Michelson-type interferometers as a cosmic probe \cite{Michelson:87}, to Bell's inequality experiments via two-photon correlations \cite{Shih:88}.
More recently, the interference of photons has proved to be an indispensable tool in quantum communications and quantum information processing, and notably in the pursuit of scalable quantum computing via quantum networked architectures of small processors (nodes) interacting via quantum channels \cite{Kimble:08, Reiserer:15}. 
Entanglement across the nodes is generated via direct entanglement transfer \cite{Cirac:97,Ritter:12} (where a mediating photon's entanglement with its emitter is transferred to an entanglement between the emitter and the recipient) or via measurement based entanglement \cite{Cabrillo:99, Feng:03, Duan:03, Simon:03, Campbell:08}.
With the former, successful entanglement is detected \it{a posteriori}\rm{} (accompanied by its collapse). The latter, albeit an inherently probabilistic protocol, is favoured for scaling as successful entanglement is heralded upon detection of readout photons \cite{Reiserer:15}.
It has led to successful elementary implementations with ions \cite{Moehring:07}, neutral atoms \cite{Hofmann:12}, defect centres in diamonds \cite{Hensen:15} and micromechanical oscillators \cite{Riedinger:18}. 
Amongst these, two-photon schemes, with flying qubits encoded in the photons' polarisations, benefit from their robustness against variations in optical phase, but the coincident detection required means high collection efficiencies are essential for high-rate entanglement generation. 

Photon production schemes using optical cavities to modify the atomic emission have permitted control of the temporal shape of the photon \cite{Vasilev:10} and have significantly improved collection efficiencies through the Purcell effect \cite{Purcell:46}, both of which play a key role in enhancing entanglement fidelities and rates \cite{Rohde:05, Goto:19}.
To this end, recent experiments have ventured toward the integration of miniature optical cavities \cite{Hunger:10, NisbetJones:11, Bransdstatter:13, Steiner:13, Gulati:17, Takahashi:20, Brekenfeld:20} with more emerging. 
The requisite cavity mirrors typically have radii of curvature of a few hundred micrometres and due to the challenges of precision manufacture of such highly curved surfaces, can often exhibit birefringence due to mirror ellipticity or anisotropic stress in the coating materials \cite{Jacob:95, Moriwaki:97, Takahashi:14, Ott:16}.
Recently, it was demonstrated that photons emitted from birefringent atom-cavity systems exhibit `polarisation oscillation' \cite{Barrett:19}, whereby the photon's polarisation varies within its wavepacket, and this effect has been harnessed as a means of enhancing extraction rates beyond the normal Purcell limit \cite{Barrett:20}. However, as we will show in this paper, in the context of polarisation-encoded remote entanglement protocols uncontrolled cavity birefringence leads to irreversible losses in entanglement fidelity and therefore must be carefully considered in such implementations.

In Sec.~\ref{formalism} we set out the formalism used to describe the two-photon measurement-based entanglement protocol, introducing the effect of a birefringent cavity at one of the nodes. In Sec. ~\ref{recovery} we discuss possible methods to reduce the impact that birefringence has on the fidelity of remote entanglement, either through photonic operations on the output or local rotations of the stationary qubit after measurement at the affected node. In Sec.~\ref{results} we present a numerical study of the loss of fidelity as a consequence of birefringence for two typical trapped ion schemes, and discuss the efficacy and technical feasibility of restoring the fidelity. Finally, we give a summary and outlook in Sec.~\ref{conclusion}.

\begin{figure}
	\includegraphics[width=0.75\textwidth]{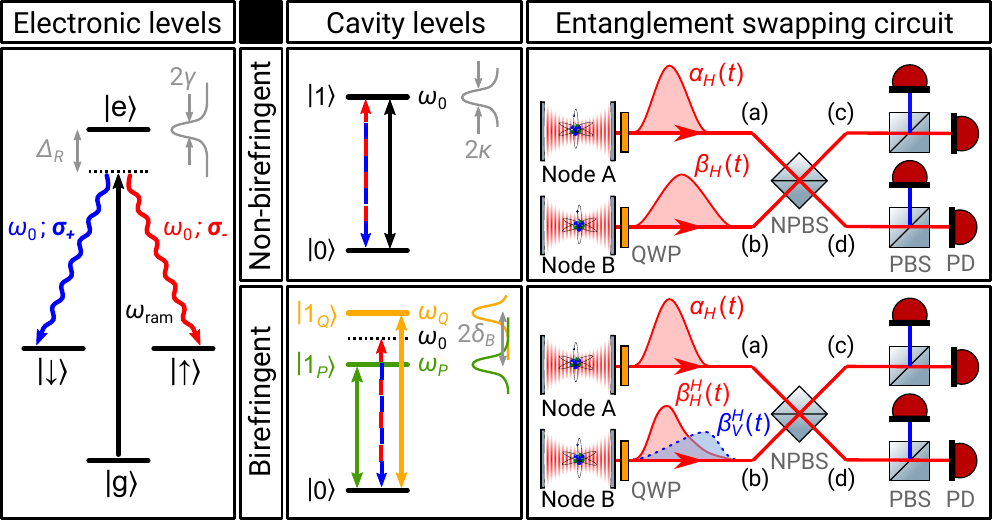}
	\caption{\label{fig:fig1} Example of a measurement-based, polarisation-encoded remote entanglement protocol between atom qubits in individual optical cavities. \textbf{(left)} Each atom is initialised in state $\ket{g}$ before laser radiation at or around frequency $\omega_{\rm{ram}}$ (detuned by $\Delta_R$ from excited state $\ket{e}$ of natural linewidth $2\gamma$) is applied to simultaneously drive coherent vSTIRAP transitions~\cite{Hennrich:00} to near-degenerate qubit states $\ket{\uparrow}$ and $\ket{\downarrow}$. These processes occur via the emission of a single photon of frequency $\sim\omega_0$ into a mode of the cavity, with its polarisation entangled to the final atomic qubit state. \textbf{(top centre)} In the case of a non-birefringent cavity with linewidth $2\kappa$ and the drive laser tuned to cavity-Raman resonance, the cavity mode at $\omega_0$ (black) couples identically and resonantly to both vSTIRAP channels (blue/red). \textbf{(top right)} To entangle the atomic qubits in two such emitter-cavity nodes, the emitted photons are directed into the input ports (a) and (b) of a non-polarising beamsplitter (NPBS), via a quarter-wave-plate (QWP) which rotates ($\sigma_-\rightarrow H$, $\sigma_+\rightarrow V$). The outputs at NPBS ports (c) and (d) are directed to polarising beamsplitters (PBS), splitting horizontal and vertical components onto photodetectors (PD); this combination of beamsplitters and detectors forms a polarisation-mode Bell-state analyser. Coincident detection of $H$ and $V$ polarised photons heralds successful entanglement-swapping and projection of the input into a two-ion-qubit Bell-state. The cavities at Nodes A and B are birefringence-free but are not necessarily identical and hence may emit temporally distinguishable photons; for simplicity we only depict the temporal profile of the $H$-polarized components ($\alpha_H$ and $\beta_H$) of the superposed states. \textbf{(bottom centre)} We next consider the case where one cavity exhibits birefringence, here for simplicity considered to be linear but arbitrarily oriented, with mode frequencies $\omega_P$ ($\omega_Q$) for polarisations $P$ and $Q$ split by $2\abs{\delta_B}$. Each vSTIRAP decay channel couples to both polarisation eigenmodes, and the emitted populations now exhibit a time-dependence, with polarisation of the intra-cavity photon wavepacket gradually rotating. \textbf{(bottom right)} Here, Node B is birefringent and so the polarisation of the photon it emits changes along its wavepacket with $\beta_H^H$($\beta_V^H$) showing the $H$ ($V$) polarised component (the superscript denotes the expected polarisation in the absence of birefringence; see Sec~\ref{intro_biref} for detailed discussion).}
\end{figure}

\section{Two-photon remote entanglement}\label{formalism}

\subsection{Non-birefringent system}		
We consider the system depicted in the top panels of Fig. \ref{fig:fig1}. An emitter-cavity system at Node A generates the balanced~\footnote{For simplicity, we consider the naturally desired balanced entangled state (which maximises entanglement generation rates) without loss of generality on the effects of birefringence. The formulation can simply be extended to an unbalanced entangled sate.} emitter-photon entangled state
\begin{equation}
    \begin{aligned}
	    \ket{\psi}_A &= \frac{1}{\sqrt{2}}\ket{\uparrow H}_A +  \ket{\downarrow V}_A
	    =  \frac{1}{\sqrt{2}}\left( \adag{}{H} \ket{\uparrow} + \adag{}{V} \ket{\downarrow} \right) \ket{0}\\
	    \ket{\psi}_B &= \frac{1}{\sqrt{2}}\ket{\uparrow H}_B +  \ket{\downarrow V}_B
	    =  \frac{1}{\sqrt{2}}\left( \bdag{}{H} \ket{\uparrow} + \bdag{}{V} \ket{\downarrow} \right) \ket{0}
    \end{aligned}
\end{equation}
where $\adag{}{x}$~($\bdag{}{x}$) is the $x$-polarised photon creation operator at Port (a)~(Port (b)) of the non-polarising beamsplitter, with $x\in\{H\textrm{(horizontal)},V \textrm{(vertical)} \}$, and $\ket{0}$ is the photon vacuum state. The photon's temporal wavepacket, which has so far not been considered, plays a key role in interference based processes and consequently on entanglement fidelities. We incorporate the photons' temporal wavepackets to the emitter-photon state by introducing the \it{photon-wavepacket creation operators} \rm \cite{Loudon:00, Takahashi:prep},
\begin{equation}
    \begin{aligned}
    \adag{}{x} \rightarrow \Adag{\alpha}{}{x} : &= \int \alpha^{*}_x(t) \adag{}{x}(t) \dd{t}
    &\qquad
    \bdag{}{x} \rightarrow \Bdag{\beta}{}{x} : &= \int \beta^{*}_x(t) \bdag{}{x}(t) \dd{t},
    \end{aligned}
    \label{eq: wavepacket_creation_definition}
\end{equation}
which hold under normalization conditions $\int \abs{\alpha_x(t)}^2 \dd{t}=\int \abs{\beta_x(t)}^2 \dd{t} = 1$ and where $ \alpha_x(t)$ ($ \beta_x(t)$) is the temporal wavepacket amplitude of the $x$-polarised emission from Node A (Node B), as shown in the right-hand panels of Fig.~\ref{fig:fig1}.

Applying the beamsplitter transformation
\begin{equation}
\mqty(a_x^{}(t) \\ b_x^{}(t)) = \frac{1}{\sqrt{2}} \mqty(1 & -1\\ 1 & 1)\mqty(c_x^{}(t) \\ d_x^{}(t)),
\end{equation}
where $c_x(t)$ and $d_x(t)$ are annihilation operators for ports (c) and (d), the wavepacket creation operators transform as
\begin{equation}
    \begin{aligned}
    \Adag{\alpha}{}{x} &= \frac{1}{\sqrt{2}}  \left(\Cdag{\alpha}{}{x}- \Ddag{\alpha}{}{x}\right)
    &\qquad \Bdag{\beta}{}{x}&= \frac{1}{\sqrt{2}}  \left(\Cdag{\beta}{}{x}+ \Ddag{\beta}{}{x}\right),
    \end{aligned}
\end{equation}
where $\Cdag{\alpha}{}{x} := \int \alpha^{*}_x(t) \cdag{}{x}(t) \dd{t}$ and $\Ddag{\alpha}{}{x} := \int \alpha^{*}_x(t) \ddag{}{x}(t) \dd{t}$. 
The output state becomes
\begin{equation}\label{eq:phit}
\begin{aligned}
\ket{\chi_{\textrm{out}}} = \frac{1}{4} \Big[
    &\left(\Cdag{\alpha}{}{H} - \Ddag{\alpha}{}{H}\right)\left(\Cdag{\beta}{}{H} + \Ddag{\beta}{}{H}\right) \ket{\uparrow\uparrow}&+ \quad&\left(\Cdag{\alpha}{}{V} - \Ddag{\alpha}{}{V}	\right)  \left(\Cdag{\beta}{}{V} + \Ddag{\beta}{}{V}	\right) \ket{\downarrow\downarrow}\\
	+&\left(\Cdag{\alpha}{}{H} \Cdag{\beta}{}{V} \ket{\uparrow \downarrow} + \Cdag{\alpha}{}{V} \Cdag{\beta}{}{H} \ket{\downarrow\uparrow} \right)&+
    \quad&\left(\Cdag{\alpha}{}{H} \Ddag{\beta}{}{V} \ket{\uparrow \downarrow} - \Ddag{\alpha}{}{V} \Cdag{\beta}{}{H} \ket{\downarrow\uparrow}\right)\\
    -&\left(\Ddag{\alpha}{}{H} \Cdag{\beta}{}{V} \ket{\uparrow \downarrow} + \Ddag{\beta}{}{H} \Cdag{\alpha}{}{V} \ket{\downarrow\uparrow}\right)&-
    \quad&\left( \Ddag{\alpha}{}{H} \Ddag{\beta}{}{V} \ket{\uparrow \downarrow} - \Ddag{\alpha}{}{V} \Ddag{\beta}{}{H} \ket{\downarrow\uparrow}\right) \Big] \ket{0}
\end{aligned}
\end{equation}
where the last four parenthetical terms correspond to the creation of $H$ and $V$ photons at ports $\{\rm{(c),(c)}\}$, $\{\rm{(c),(d)}\}$, $\{\rm{(d),(c)}\}$ and $\{\rm{(d),(d)}\}$ respectively; the successful detection of each of these outcomes provides a unique herald. The projected atomic state upon detection of $H$ and $V$ photons at Port (c) at times $t_H$ and $t_V$ is
\begin{equation}
\begin{aligned}
\ket{\psi_{cc}(t_H,t_V)} &=\bra{0}\cann{H}{t_H} \cann{V}{t_V} \ket{\chi_{\textrm{out}}} \\ &= \frac{1}{4} \bra{0}\cann{H}{t_H} \cann{V}{t_V}  \Big[\Cdag{\alpha}{}{H} \Cdag{\beta}{}{V} \ket{\uparrow \downarrow} +  \Cdag{\alpha}{}{V} \Cdag{\beta}{}{H} \ket{\downarrow \uparrow } \Big] \ket{0} \\
& = \frac{1}{4} \Big[ \iint \alpha^*_H(s_H) \delta(t_H-s_H) \beta^*_V(s_V) \delta(t_V-s_V) \dd{s_H}\dd{s_V} \ket{\uparrow \downarrow} \quad \\
& + \iint \alpha^*_V(s_V) \delta(t_V-s_V) \beta^*_H(s_H) \delta(t_H-s_H) \dd{s_H}\dd{s_V}\ket{\downarrow \uparrow} \Big] \\
& = \frac{1}{4} \left(\alpha^*_H(t_H) \beta^*_V(t_V) \ket{\uparrow\downarrow }+ \alpha^*_V(t_V) \beta^*_H(t_H) \ket{\downarrow \uparrow}\right), 
\end{aligned}
\end{equation}
where we have used $[\cann{x}{t_x},\cdag{}{x}(s_x)] = \delta(t_x-s_x)$.

The fidelity of this state with the Bell state $\ket{\Psi_{\textrm{Bell}}^+} = \frac{1}{\sqrt{2}}(\ket{\uparrow \downarrow} +\ket{\downarrow\uparrow})$ is given by
\begin{equation}\label{eq:FidelityNoBiref}
	\mathcal{F}_{cc}(t_H,t_V)= \frac{1}{\Tr{\ket{\psi_{cc}}\bra{\psi_{cc}}}}\abs{\braket{\Psi_{\textrm{Bell}}^+}{\psi_{cc}}}^2 
	= \frac{1}{2}\left( 1+\frac{\alpha^*_H(t_H) \beta_H(t_H)\beta^*_V(t_V) \alpha_V(t_V) + \text{c.c.}}{\abs{\alpha_H(t_H) \beta_V(t_V)}^2 + \abs{\alpha_V(t_V)\beta_H(t_H)}^2}\right). 
\end{equation}
Integrating over all detection events, where $p_{cc}(t_H, t_V)$ is the unity-normalised probability density for a $\{\rm{(c),(c)}\}$ herald occurring at times $t_H$ and $t_V$, we find the fidelity to be
\begin{equation}
    \begin{aligned}
	\mathcal{F}_{cc} &= \int{\int{\mathcal{F}_{cc}(t_H,t_V) p_{cc}(t_H,t_V))\dd{t_H}}\dd{t_V}}\\
	&=\frac{1}{2} \int{\int{\mathcal{F}_{cc}(t_H,t_V)\left(\abs{\alpha_H(t_H) \beta_V(t_V)}^2 + \abs{\alpha_V(t_V)\beta_H(t_H)}^2\right)\dd{t_H}}\dd{t_V}}\\
     &= \frac{1}{2} \left(1 + \Re{\int \dd{t_H}\alpha^*_H(t_H) \beta_H(t_H) \int \dd{t_V}\beta^*_V(t_V) \alpha_V(t_V)}\right).
	\end{aligned}
\end{equation}
When the emission wavepackets from Nodes A and B are identical, the fidelity with the Bell state becomes unity. A corresponding relation holds for fidelity $\mathcal{F}_{dd}$ with $\ket{\Psi^{+}_{\rm{Bell}}}$, while $\mathcal{F}_{cd}$ and $\mathcal{F}_{dc}$ are unity for overlap with $\ket{\Psi^-_{\rm{Bell}}}=\frac{1}{\sqrt{2}}(\ket{\uparrow \downarrow} -\ket{\downarrow\uparrow})$.

\subsection{Introducing birefringence}\label{intro_biref}

We now consider the case where the cavity at Node B exhibits birefringence, where the cavity resonance $\omega_0$ is split into two non-degenerate polarisation eigenmodes at frequencies $\omega_Q$ and $\omega_P$, with a birefringent splitting $2\delta_B = \abs{\omega_Q-\omega_P}$. It has been shown that the polarisation of a photon emitted from a birefringent cavity will generally exhibit a time dependence~\cite{Barrett:19}. The small difference in effective path length for the polarisation eigenmodes of the cavity leads to a relative phase difference that accumulates over the many round-trips within the cavity. Unless the emitted photon's polarisation matches one of these eigenmodes, this results in a gradual rotation of its polarisation state. If the path difference accumulates faster than the cavity decay rate, a time-dependent polarisation will be observed, as depicted in the bottom-right panel of Fig.~\ref{fig:fig1}.

The birefringence transforms the wavepacket $\beta_H(t)$  to a superposition of an $H$ component with wavepacket $\beta^H_H(t)$, and a $V$ component with wavepacket $\beta^H_V(t)$. The standard (non-birefringent) photon-wavepacket creation operator (see Eq. \ref{eq: wavepacket_creation_definition}) can thus be substituted with the following for Port (b), which is coupled to the birefringent node:
\begin{align}
	\Bdag{\beta}{}{H} &\rightarrow n^{\beta,H}_H\Bdag{\beta}{,H}{H} +n^{\beta,H}_V \Bdag{\beta}{,H}{V},
	&\qquad
	\Bdag{\beta}{}{V} &\rightarrow  n^{\beta,V}_V\Bdag{\beta}{,V}{V} + n^{\beta,V}_H\Bdag{\beta}{,V}{H}.
\end{align}
Here, $\Bdag{\beta}{,y}{x}$ are wavepacket creation operators, with $\beta$ indicating the temporal wavepacket amplitude of a photon emitted by Node B, $y$ indicating the `nominal' photon polarisation and $x$ the polarisation of the photon created such that, e.g., $\Bdag{\beta}{,H}{V} = \int \beta^{H*}_V(t)\bdag{}{V}(t) \dd{t}$. The normalisation coefficients  $n^{\beta,y}_x$ ensure that $\Bdag{\beta}{,y}{x}$ obey the $\int{\abs{\beta^{y*}_x(t)}^2\dd{t}}=1$ normalisation condition of the wavepacket creation operator formalism, while the overall integrated probability of a single photon being produced across the two polarisation modes remains unity. This transformation makes no assumptions about the relative decay rate for each cavity eigenmode nor about potential back action of the emission onto the emitter-cavity dynamics, and in general the $\beta^{y*}_x(t)$ are each unique and not trivially related. The overall action of the birefringence is unitary on the combined ion-qubit/photon-qubit Hilbert space of Node B, however its action on each qubit subspace alone is generally not. This has important implications for the feasibility of active correction of the polarisation oscillations, which we will discuss in Sec.~\ref{photon_corr}.

For simplicity, from this point on we will use non-normalised wavepacket creation operators, defined as
\begin{equation}\label{eq:Btilde}
\tBdag{\beta}{,y}{x} = n_x^{\beta,y} B^{\beta,y}_x=\int{\tilde{\beta}_x^{y*}(t)\bdag{}{x}(t)\dd{t}}
\end{equation}
The resulting state at Node $B$ thus becomes
\begin{equation}\label{eq:NodeBBiref}
	\ket{\psi'}_B = \frac{1}{\sqrt{2}}\Big(\left(\tBdag{\beta}{,H}{H} + \tBdag{\beta}{,H}{V}\right)\ket{\uparrow}  + \left(\tBdag{\beta}{,V}{H} + \tBdag{\beta}{,V}{V}\right)\ket{\downarrow}\Big).
	\end{equation}
$\tBdag{\beta}{,H}{x}$ and  $\tBdag{\beta}{,V}{x}$ transform similarly to $\Bdag{\beta}{}{x}$ at the beamsplitter with $\tBdag{\beta}{,y}{x} = \frac{1}{\sqrt{2}}  \left(\tCdag{\beta}{,y}{x} + \tDdag{\beta}{,y}{x}\right)$.

If we consider again the coincident detection of one $H$- and one $V$-polarised photon at Port (c), the relevant terms in the expansion of $\ket{\chi_{\textrm{out}}'} = \ket{\psi}_A\otimes\ket{\psi'}_B$ are
\begin{align*}
	\Cdag{\alpha}{}{H}\tCdag{\beta}{,H}{V}\ket{\uparrow\uparrow}\rm{, }\qquad \Cdag{\alpha}{}{H}\tCdag{\beta}{,V}{V}\ket{\uparrow\downarrow}\rm{, }\qquad \Cdag{\alpha}{}{V}\tCdag{\beta}{,H}{H}\ket{\downarrow\uparrow}\rm{, }\qquad\rm{and }\qquad \Cdag{\alpha}{}{V}\tCdag{\beta}{,V}{H}\ket{\downarrow\downarrow}.
\end{align*}
We find the projected atomic state after the detection of $H$ and $V$ photons at Port (c) at times $t_H$ and $t_V$ to be:
\begin{equation}\label{eq:BirefOut1}
	\begin{aligned}
	\ket{\psi_{cc}'}&= \bra{0} \cann{H}{t_H}\cann{V}{t_V}\ket{\chi_{\textrm{out}}'}\\
	&= \frac{1}{4} \bra{0}\cann{H}{t_H}\cann{V}{t_V} \Big[ \Cdag{\alpha}{}{H}\ket{\uparrow}\Big(\tCdag{\beta}{,H}{V}\ket{\uparrow}  +  \tCdag{\beta}{,V}{V}\ket{\downarrow}\Big) + \Cdag{\alpha}{}{V}\ket{\downarrow}\left(\tCdag{\beta}{,H}{H}\ket{\uparrow} + \tCdag{\beta}{,V}{H}\ket{\downarrow}\right) \Big]\ket{0}\\
	&= \frac{1}{4} \Big[ \alpha^{*}_H(t_H) \ket{\uparrow} \left( \tilde{\beta}^{H*}_V(t_V) \ket{\uparrow} + \tilde{\beta}^{V*}_V(t_V) \ket{ \downarrow}  \right) + \alpha^{*}_V(t_V) \ket{\downarrow} \left( \tilde{\beta}^{H*}_H(t_H) \ket{\uparrow} + \tilde{\beta}^{V*}_H(t_H) \ket{ \downarrow}  \right) \Big]. 
	\end{aligned}
\end{equation}
We note that $\ket{\psi_{cc}'}$ appears similar to $\ket{\Psi_{\rm{Bell}}^+}$, but with the terms transformed as $\{\ket{\uparrow\downarrow},\ket{\downarrow\uparrow}\}\rightarrow\{\ket{\uparrow\downarrow'},\ket{\downarrow\uparrow'}\}$, where
\begin{equation}\label{eq:primed_defs}
\ket{\downarrow'} = \alpha^{*}_H(t_H)\left(\tilde{\beta}^{H*}_V(t_V) \ket{\uparrow} + \tilde{\beta}^{V*}_V(t_V) \ket{ \downarrow}  \right) \hspace{10pt}\rm{and}\hspace{10pt}
\ket{\uparrow'}=\alpha^{*}_V(t_V)\left( \tilde{\beta}^{H*}_H(t_H) \ket{\uparrow} + \tilde{\beta}^{V*}_H(t_H) \ket{ \downarrow}  \right).
\end{equation}
The fidelity with the Bell state $\ket{\Psi_{\textrm{Bell}}^+} = \frac{1}{\sqrt{2}}(\ket{\uparrow \downarrow} +\ket{\downarrow\uparrow})$ is given by
\begin{equation}
\begin{aligned}
\mathcal{F}&= \frac{1}{\Tr{\rho}}\abs{\braket{\Psi_{\textrm{Bell}}^+}{\psi_{cc}'}}^2 \\
		&= \frac{1}{2}\Big[1 + \frac{ \Re{\alpha^{*}_H(t_H) \tilde{\beta}^H_H(t_H)\alpha_V(t_V) \tilde{\beta}^{V*}_V(t_V)}}{8\Tr{\rho}}
		-\frac{\abs{\alpha_H(t_H) \tilde{\beta}^H_V(t_V)}^2 + \abs{\alpha_V(t_V) \tilde{\beta}^V_H(t_H)}^2}{16\Tr{\rho}}  \Big]
	\end{aligned}
	\label{eq:single event fidelity}
\end{equation}
where $\rho = \ket{\psi_{cc}'}\bra{\psi_{cc}'}$. This resembles the non-birefringent case~\eqref{eq:FidelityNoBiref}, but fidelity now decreases with increasing birefringence due to the general growth in amplitudes of `rotated' polarisation components $\tilde{\beta}^V_H(t_H)$ and $\tilde{\beta}^H_V(t_V)$.

\section{Recovery of Fidelity}\label{recovery}

In Sec.~\ref{intro_biref}, we have seen that cavity birefringence in a network node can lead to a temporal variation of the output photon polarisation. This effect reduces the fidelity of the atomic Bell state produced via entanglement swapping, upon the coincident (Bell-basis) detection of the birefringent cavity photon and that extracted from another node. We now consider several approaches to reducing this fidelity loss, either by attempting to `undo' the polarisation oscillation directly, or by atomic state post-selection or post-correction, conditioned on the photon arrival times.

\begin{figure}
    \centering
    \includegraphics[width=\textwidth,trim={1cm 6.5cm 6cm 6.5cm},clip]{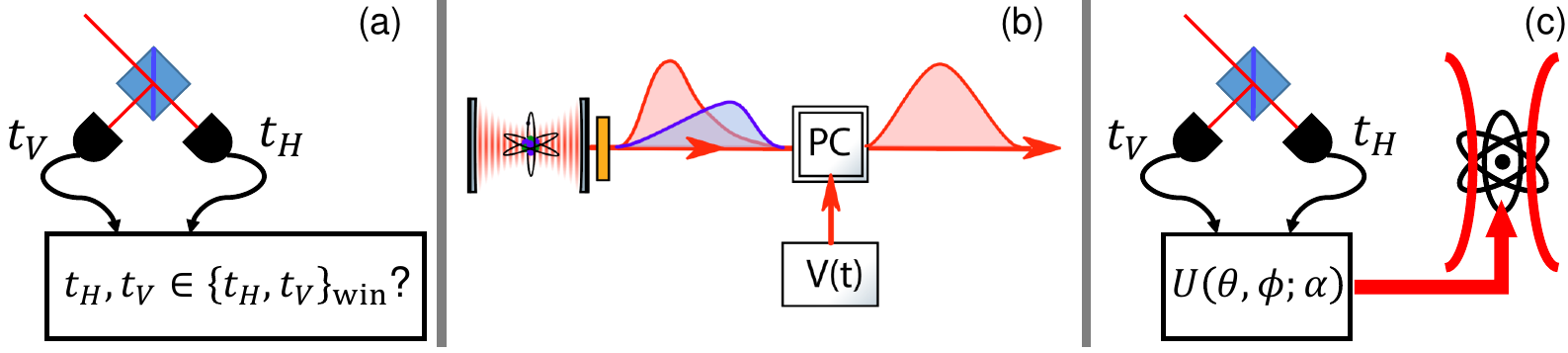}
\caption{We consider three schemes for improving entanglement fidelity in the presence of cavity birefringence: \textbf{(a)} Windowing: Events are post-selected on the basis that the herald photon arrival times $t_H, t_V$ fall within a pre-defined window region $\{t_H, t_V\}_\text{win}$; \textbf{(b)} Photonic correction: the output from the cavity in each node is directed through a dedicated Pockell's cell (PC), to which a voltage waveform $V(t)$ is applied to achieve a time-dependent polarisation rotation that approximately reverses the cavity-induced polarisation oscillation; \textbf{(c)} Spin-qubit correction: An arbitrary unitary transformation is applied to a single atom post-herald, with the particular rotation axis (defined by spherical coordinates $\theta, \phi$) and rotation angle ($\alpha$) determined by $t_H, t_V$.}
    \label{fig:recovery_schemes}
\end{figure}

\subsection{Pre-herald photonic qubit correction}\label{photon_corr}

We first consider whether the effects of birefringence, namely the transformation of the photonic wavepackets, can be restored before interference on the beamsplitter by using a time-dependent linear optical device between Node B and Port (b) to invert this transformation (Fig.~\ref{fig:recovery_schemes}(b)). Two distinct categories of transformation occur - a time-dependent rotation of the polarisation along the wavepacket, and a distortion of the total wavepacket amplitude due to gradual decoupling of the rotated cavity field from the atom. We will discuss these two effects in turn.

The first impact of cavity birefringence is to introduce a time-dependence to the photon polarisation correlated with each atomic state. In the case of zero birefringence, the node state $\ket{\uparrow}$ is always correlated with photon polarisation $H$, and $\ket{\downarrow}$ with $V$; a birefringent node produces a much more general state with spin-polarisation correlations that evolve in time, as given in Eqs.~(\ref{eq:Btilde}) and (\ref{eq:NodeBBiref}). To recover the ideal joint atom-photon state, we require a transformation that rotates the component correlated with $\ket{\uparrow}$ to $H$, and that correlated with $\ket{\downarrow}$ to $V$, at all times $t$:
\begin{equation}
    \begin{aligned}
        \tilde{\beta}_H^{H*}(t)\bdag{}{H}(t)+\tilde{\beta}_V^{H*}(t)\bdag{}{V}(t)\rightarrow\hat{\tilde{\beta}}_H^*(t)\bdag{}{H}(t)
        &\qquad \text{and}
        &\qquad
        \tilde{\beta}_H^{V*}(t)\bdag{}{H}(t)+\tilde{\beta}_V^{V*}(t)\bdag{}{V}(t)\rightarrow\hat{\tilde{\beta}}_V^*(t)\bdag{}{V}(t)
    \end{aligned}
\end{equation}
However, this polarisation transformation is only possible via a unitary operation if the Jones vectors for the photonic components correlated with $\ket{\uparrow}$ and $\ket{\downarrow}$ remain orthogonal to one another at all times; explicitly
\begin{equation}
    \begin{aligned}
        \left(\mqty{\tilde{\beta}_H^{H}(t) & \tilde{\beta}_V^{H}(t)}\right) \bullet \left(\mqty{\tilde{\beta}_H^{V*}(t) \\ \tilde{\beta}_V^{V*}(t)}\right)=0; &\qquad \forall t>0.
    \end{aligned}
\end{equation}
This condition only holds for certain special cases, such as the case of complete symmetry of the emitter-cavity system with respect to the permutation of $\{\ket{H \uparrow},\ket{V \downarrow}\}$. For cases and times when this orthogonality is violated, no unitary operation on the emitted photon will be able to simultaneously invert the polarisation transformation of both $\ket{\uparrow}$-correlated and $\ket{\downarrow}$-correlated components. Nonetheless, in practice (e.g. for the examples discussed in Sec.~\ref{results}) the best possible unitary operation can remedy the polarisation oscillations of the cavity emission at most times effectively.

The second impact of birefringence is to modify the time dependence of the overall wavepacket amplitude. The shape of the photon wavepacket is governed by the coupled interactions of laser, atom and cavity, and when the polarisation of the intra-cavity field rotates due to birefringence, the cavity-field/atom coupling is modified, changing the dynamics of the system~\cite{Barrett:20}. The result is that even if the spin-polarisation entanglement can be fully restored, the wavepackets emitted from nodes with differing degrees of birefringence will still differ, because a unitary photonic correction cannot change the overall amplitude of the wavepacket at any given time. This reduces the overlap of the photonic wavepackets from the two nodes and, for the cases we consider, this distinguishability provides the principal limit to the fidelity that may be recovered via photonic correction.

For the corrections presented in Sec.~\ref{results} we model the effect of a single variable retarder such as a Pockels cell with retardance varying linearly in time. The cell orientation, initial retardance and ramp rate are optimised such that in the basis $\{\ket{H\uparrow},\,\ket{V\uparrow},\,\ket{H\downarrow},\,\ket{V\downarrow}\}$ the integrated amplitudes of the latter two `anti-correlated' terms are minimised. This is not the highest performance photonic correction scheme possible, but has the considerable advantage that the photonic correction could be calibrated via analysis of the emission of the corrected node alone, meaning that the simple three-parameter optimisation need only be done once per network node, and requires only single photon detection events.

Higher fidelities can be achieved with more complete photonic correction methods, such as using more complex temporal profiles for the retardance or multiple Pockels cells at different orientations. One can also choose to optimise on the two-node fidelity for higher performance, but the requirement of two-photon coincidences during calibration, which would need to be repeated for every pairwise combination of nodes in the network, drastically increases the calibration time required. In our simulations, the relative improvement seen from using these more complicated photonic corrections versus the approach selected was generally negligible.

\subsection{Post-herald atomic qubit correction}\label{spin_corr}

\subsubsection{Windowed post-selection}\label{spin_corr_win}
The probabilistic remote entanglement scheme under consideration consists of repeated attempts that terminate upon a suitable `$H$--$V$' two-photon herald, after which the atoms in the nodes are considered to be entangled. By applying additional criteria to the acceptance of a herald (Fig.~\ref{fig:recovery_schemes}(a)) it is possible to improve the fidelity of the Bell-pair production process in the presence of cavity birefringence, at the expense of success rate.

Eq.~\ref{eq:single event fidelity} shows that the fidelity of an event is a function of detection times $t_H$ and $t_V$. By deeming a trial successful only if the expected fidelity at those detection times exceeds a certain threshold, the average fidelity of successful trials will improve. The efficacy of this windowed post-selection approach depends upon the accuracy with which the detection times can be measured, and the correspondence between the model of $\mathcal{F}(t_H,t_V)$ used and reality.

Windowing may also be used in conjunction with the other correction methods in this section; in these cases the fidelities expected after application of those corrections are used to determine suitable windowing regions.

\subsubsection{Spin-qubit rotation}\label{spin_corr_rot}

The fidelity of the Bell state produced can also be partially recovered via local operations on the two atomic qubits post-detection (Fig.~\ref{fig:recovery_schemes}(c)). Because $\ket{\uparrow'}$ and $\ket{\downarrow'}$ are governed by independent photon wavepacket functions, $\tilde{\beta}_H^{x}(t_H)$ and $\tilde{\beta}_V^{x}(t_V)$, cavity birefringence causes more than a trivial unitary rotation in bases from $\ket{\uparrow}$ and $\ket{\downarrow}$, even when $t_H=t_V$. This means that the two-qubit state $\ket{\psi_{cc}'}$ is not generally fully entangled for all $t_H$ and $t_V$, and no local operations on the qubits post-herald can completely restore the fidelity.

However, significant improvement in fidelity is nonetheless achievable via local unitary rotations. Because the \textit{target} state is fully entangled, the correction need only be applied to one qubit, and the optimal unitary is uniquely determinable (see App.~\ref{app:unique and single node correction}). However, the correction unitary depends upon the detection times $t_H$ and $t_V$, and in practice these corrections would be arduous to implement as they would have to be calibrated for all pairs of nodes, heralds and combinations of photon arrival times.

\newpage
\section{Results}\label{results}

We now numerically analyse the impact of cavity birefringence on the remote entanglement protocol. Our analysis up to this point has considered generic emitter-photon systems, but we now focus on two distinct experimentally-relevant trapped ion system configurations, the detailed dynamics of which differ markedly. However the broad results and conclusions are similar for the two cases, and comparable behaviour would be observed for neutral atom systems.

\subsection{Spin-photon entanglement schemes}

\subsubsection*{$\sigma_-/\sigma_+$ encoding}

We first consider a $\sigma_-/\sigma_+$ photon encoding scheme, shown in the upper row of Fig.~\ref{fig:sr_yb_schemes}. This encoding can be engineered in a range of systems, but naturally lends itself to an ion with simple hyperfine structure, such as the case of $^{171}\text{Yb}^+$, shown here. The symmetric Clebsch-Gordan coefficients ensure equal coupling on both decay channels, and the cavity axis is aligned parallel to the magnetic field, meaning that $\pi$-polarised decays are unsupported by the cavity mode. This means that the system is overwhelmingly likely to decay on the target channels without additional frequency-selective suppression of the third `spectator' decay, and can thus be operated close to Zeeman degeneracy (for example with qubit splitting $\Delta_z = 0.5\kappa$ and driven by a single Raman laser frequency. Upon emission from the cavity, a quarter-wave retarder transforms the polarisation $\sigma_-/\sigma_+\rightarrow H/V$ to permit compatibility with the same Bell state analyser.

For our simulations of $\sigma_-/\sigma_+$ encoding, we use $g_{\uparrow}=g_{\downarrow}=\kappa$ and $\gamma = 0.6\kappa$ with the atomic states separated by energy $0.5 \kappa$, where $2\kappa$ is the cavity linewidth and $g_{\uparrow}$ ($g_{\downarrow}$) is the cavity coupling strength between the excited state $\ket{e}$ and qubit state $\ket{\uparrow}$ ($\ket{\downarrow}$). Note that the selection of $g_{\uparrow}, g_{\downarrow}\sim\kappa$ is near-ideal for optimal extraction efficiency \cite{Goto:19}, and the choice $\gamma = 0.6g$ is loosely based on currently achievable desired coupling regimes \cite{Takahashi:20}.

\subsubsection*{$H/V$ encoding}

\begin{figure}
	\includegraphics[width=0.66\textwidth]{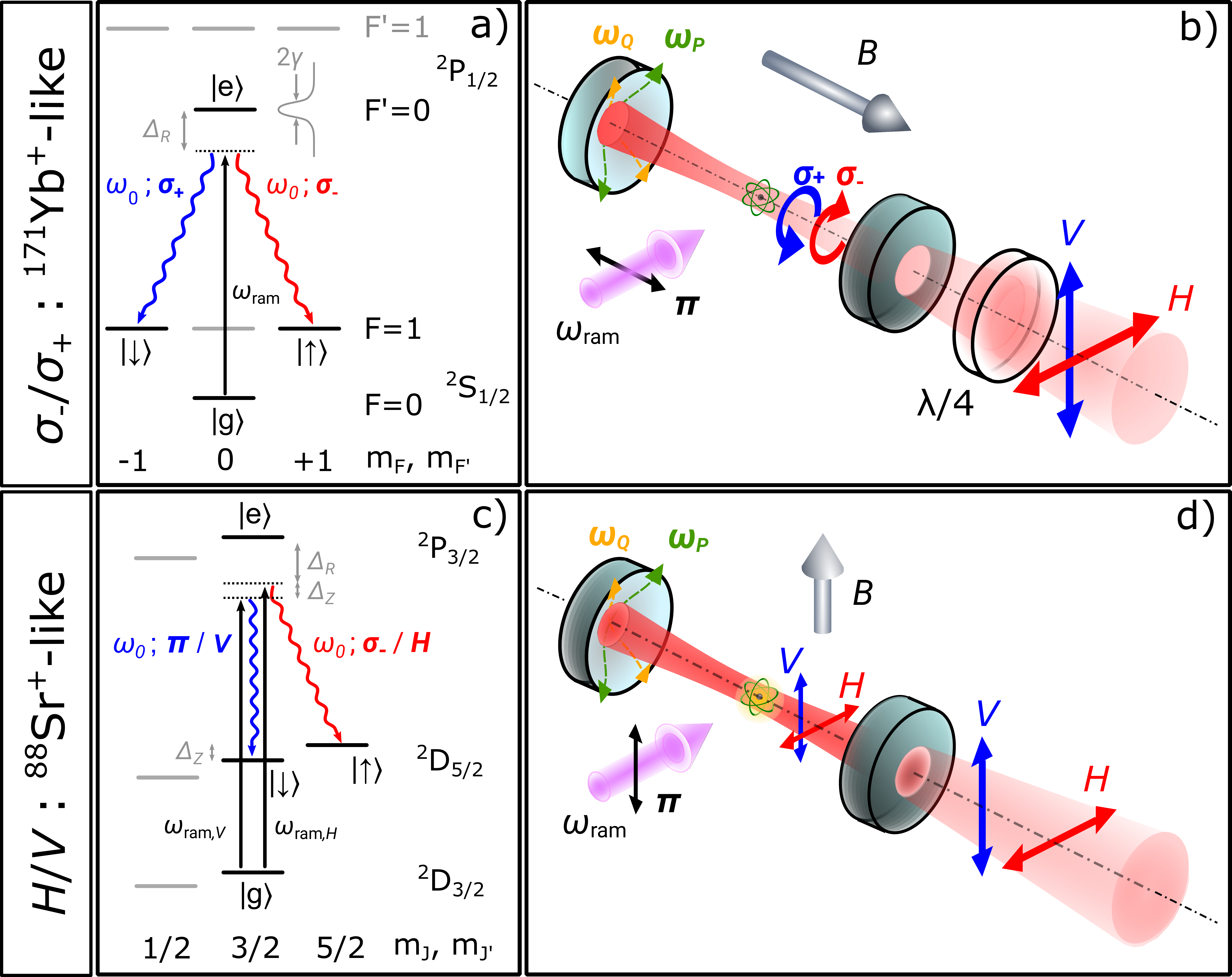}
	\caption{We consider two classes of ion-photon entanglement schemes, which lead to qualitatively different behaviour, for both birefringent and non-birefringent systems. \textbf{(a)} Typical level scheme for $\sigma_-/\sigma_+$ photon encoding, suitable for implementation in isotopes with simple hyperfine structure e.g. $^{171}\text{Yb}^+$ operated close to Zeeman degeneracy ($B\approx5\text{G}$). The ion is prepared in ground state $\ket{g}$ before a single Raman laser at $\omega_\text{ram}$ is applied to drive simultaneous symmetric vSTIRAP transfers via excited state $\ket{e}$ (with lifetime $(2\gamma)^{-1}$) to spin qubit states $\ket{\uparrow}/\ket{\downarrow}$, accompanied by the emission of $\sigma_-/\sigma_+$ polarised photons. The cavity mode, in the absence of birefringence, is set to frequency $\omega_0$, which is detuned by $\Delta_R$ from the qubit-excited state transition frequency $\omega_{e\uparrow}$.	
	\textbf{(b)} The cavity is arranged with optical axis parallel to the B-field, where it provides maximum Purcell enhancement on $\sigma_-/\sigma_+$ decays. For the case of birefringent nodes, the cavity resonance $\omega_0$ is split into two non-degenerate polarisation eigenmodes at frequencies $\omega_Q$ and $\omega_P$. Photons transmitted by the cavity output mirror pass through a quarter-wave plate, transforming $\sigma_-/\sigma_+\rightarrow H/V$, before transmission into the Bell-state analyser or wider quantum network.
	\textbf{(c)} Typical level scheme to $H/V$ photon encoding, suitable for implementation in e.g. $^{40}\text{Ca}^+$ or $^{88}\text{Sr}^+$. For such isotopes, lacking convenient hyperfine structure, the most natural vSTIRAP scheme involves preparation in a ground or metastable state, followed by vSTIRAP transfer to other metastable levels via dipole-coupled excited states. With a single valence electron and no nuclear spin, we lack the symmetry needed for balanced decays on $\sigma_-/\sigma_+$ channels, and decay rates are instead most balanced when transferring to a qubit encoded in the most extreme states of the lower manifold, here $\ket{\uparrow}=D_{5/2,+5/2}$ and $\ket{\downarrow}=D_{5/2,+3/2}$, accompanied by the emission of $\sigma_-$ and $\pi$ polarised photons. In this case, decay is also possible via the $\sigma_+$ channel to `spectator' level $D_{5/2,+1/2}$, reducing the fidelity of the final state. This process is suppressed by operating with substantial Zeeman splitting, $\Delta_Z$ (typically $5\kappa$), which in turn requires that the two vSTIRAP $\lambda$-systems are addressed by dedicated frequency components $\omega_{\text{ram},H}/\omega_{\text{ram},V}$ of a bichromatic drve, each tuned to Raman resonance with the corresponding qubit state via cavity mode frequency $\omega_0$. \textbf{(d)} By aligning the cavity axis orthogonal to the magnetic field, the $\sigma_-/\pi$ decay channels are projected directly to H/V without a waveplate, while the difference in their angular emission patterns partially cancels the strength imbalance of their Clebsch-Gordan coefficients.}
	\label{fig:sr_yb_schemes}
\end{figure}

The second scheme we consider directly encodes the photon in an $H/V$ basis via projection of $\sigma_-$ and $\pi$ decays into a cavity with axis orthogonal to the magnetic field, shown in the lower row of Fig.~\ref{fig:sr_yb_schemes}. While the Clebsch-Gordan coefficients for each channel are never exactly equal, when also accounting for the angular dependence of the atom-field coupling the difference can be made small by working with `stretch' transitions at an extremum of a Zeeman manifold, such as in the $^{40}\text{Ca}^+$-- or $^{88}\text{Sr}^+$--like case shown here. Because $\sigma_+$ decays project to the same cavity photon polarisation as $\sigma_-$, the system must be operated far from Zeeman degeneracy (typically $\Delta_z\approx5\kappa$) to ensure resonant suppression of this unwanted decay channel. This in turn means that the two qubit channels cannot be driven with a single Raman laser frequency, and a bichromatic drive is required, leading to a more complex output photon wavepacket as will be seen later. However, no further ex-cavity polarisation optics are necessary.

For our simulations of $H/V$ encoding, the Clebsch-Gordan coefficients of the cavity decays are not equal, and the projection of the $\sigma_+$ atomic emission onto the $h$-polarised cavity mode is not unity, leading to $g_{\uparrow}=\sqrt{1/3}g_0$ and $g_{\downarrow} = -\sqrt{4/15}g_0$ with $g_0$ chosen such that $g_{\uparrow}^2+g_{\downarrow}^2 = 2\kappa^2$. The two qubit states are separated in energy by $5\kappa$ so that off-resonant decays to non-qubit states are strongly suppressed. $\kappa$ and $\gamma$ remain unchanged from the $\sigma_-/\sigma_+$ case.

\subsection{Simulation of raw and corrected remote entanglement performance}
We now simulate the remote entanglement protocol for the two schemes above, using the methods described in App.~\ref{app:sim_method} to calculate the photon temporal wavepackets at each node output for both the ideal and birefringent nodes, and thus the state $\ket{\chi_\text{out}}$ of the whole system post-NPBS. Following Eq.~\ref{eq:BirefOut1}, we then find the two-qubit states produced as a result of detection events at times $t_H$ and $t_V$, and the fidelity of these states with the target Bell state, as given in Eq.~\ref{eq:single event fidelity}. Finally, we consider how each of the mitigation strategies described in Sec.~\ref{recovery} affects the fidelity.

We begin by presenting a selection of concrete examples highlighting the effect of cavity birefringence on the photonic wavepackets and the resultant remote entanglement fidelity as a function of the detection times, before presenting the corrections that should be applied to restore the fidelity. We then investigate how the achievable fidelity varies with birefringence orientation and magnitude. We will find that the choice of encoding, the birefringence magnitude $\abs{\delta_B}$, and its orientation $\arg(\delta_B)$ are all important determinants of the impact of birefringence upon the achievable fidelity pre- and post-correction, as well as upon the difficulty of applying the necessary corrections. 

\subsubsection*{Photon production and correction dynamics}
\color{black}
The first results study the effect of birefringence on the output wavepackets of nodes and the corrections that should be applied for a range of specific cases. Considering the $\sigma_-/\sigma_+$ and $H/V$ schemes in turn, in each case one node is assumed to be ideal while the other experiences a cavity birefringence of magnitude $\abs{\delta_B}$ whose principal axes are at an angle of $\pi/4$ to the photon emission axes on the Poincare sphere (for more details, see App.~\ref{app:sim_method}). 
This intermediate angle is chosen to give typical dynamics combining both `polarisation-oscillation'-type rotations between cavity photon mode populations, and a mode-dependent phase shift.

\begin{figure}[]
    \centering
    \includegraphics[width=1.0\textwidth, trim=2.5cm 1.75cm 3.5cm 0.05cm, clip]{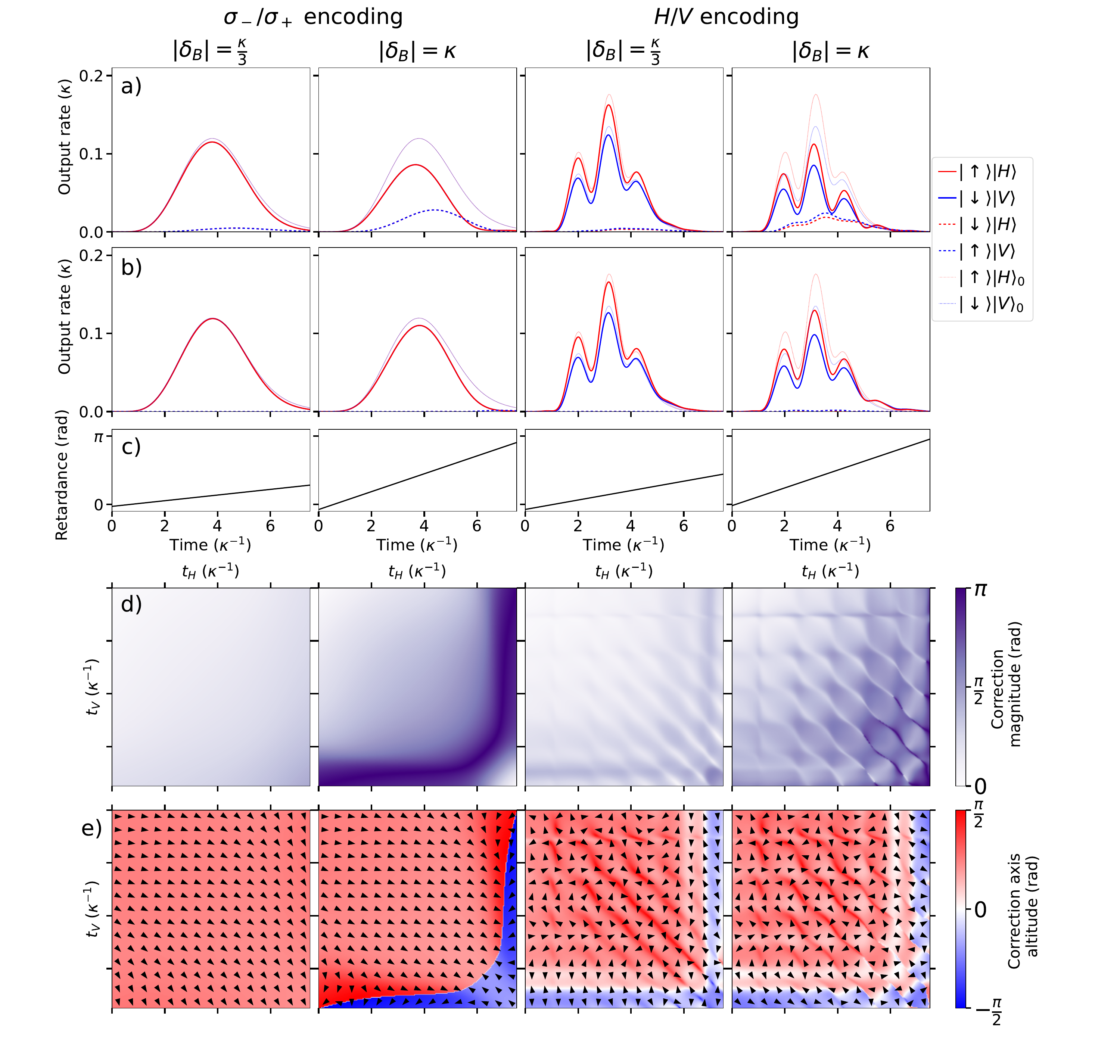}
    \caption{Examples of node output wavepackets and pre-herald photonic or post-herald spin corrections for the $\sigma_-/\sigma_+$ (Columns 1 \& 2) and $H/V$ (Columns 3 \& 4) systems in the presence of weaker or stronger birefringence (left and right of the two columns respectively). In each case, one node is ideal (zero birefringence) while the other exhibits birefringence of labelled magnitude $\abs{\delta_B}$, applied at an angle of $\pi/4$ to the atomic emission axes on the Poincare sphere. Rows \textbf{(a)-(e)} compare features of the wavepackets and the structure of the possible correction operations. \textbf{(a)} Output of the birefringent node before any correction is applied. The solid red and blue lines show the output rate the correlated atom-photon states that would be expected in the absence of birefringence; the heavy dashed lines represent the output rate of the opposite correlation. The polarisation of the cavity photon varies during the production process, with the `rotated' polarisation components more prevelant for higher birefringence. Note the feint dotted lines indicating the output of the ideal node in the absence of birefringence.  \textbf{(b)} Output of the birefringent node after photonic correction; note the spin-polarisation correlation is near-perfectly restored but the wavepacket still deviates from the non-birefringent case. \textbf{(c)} Linearly varying retardance as a function of time for a single Pockels cell, with retardance ramp-rate and fast-axis orientation chosen for optimal restoration of the expected photon correlations (see Sec.~\ref{photon_corr}). \textbf{(d)} Magnitude of optimal single spin-qubit unitary correction (see Sec.~\ref{spin_corr}), applied to the ion in the birefringent node after a $\{(\rm{c}),(\rm{c})\}$ herald leaves the system in the state given by Eq.~\ref{eq:BirefOut1}. \textbf{(e)} Corresponding orientation of the unitary rotation axis on the Bloch sphere, with the arrows and heatmap indicating the azimuthal and altitudeinal angles respectively.}
    \label{fig:hyper_plot_wavefunction_and_correction}
\end{figure}

\begin{figure}[]
    \centering
    \includegraphics[width=1.0\textwidth, trim=3cm 3cm 3cm 0.05cm, clip]{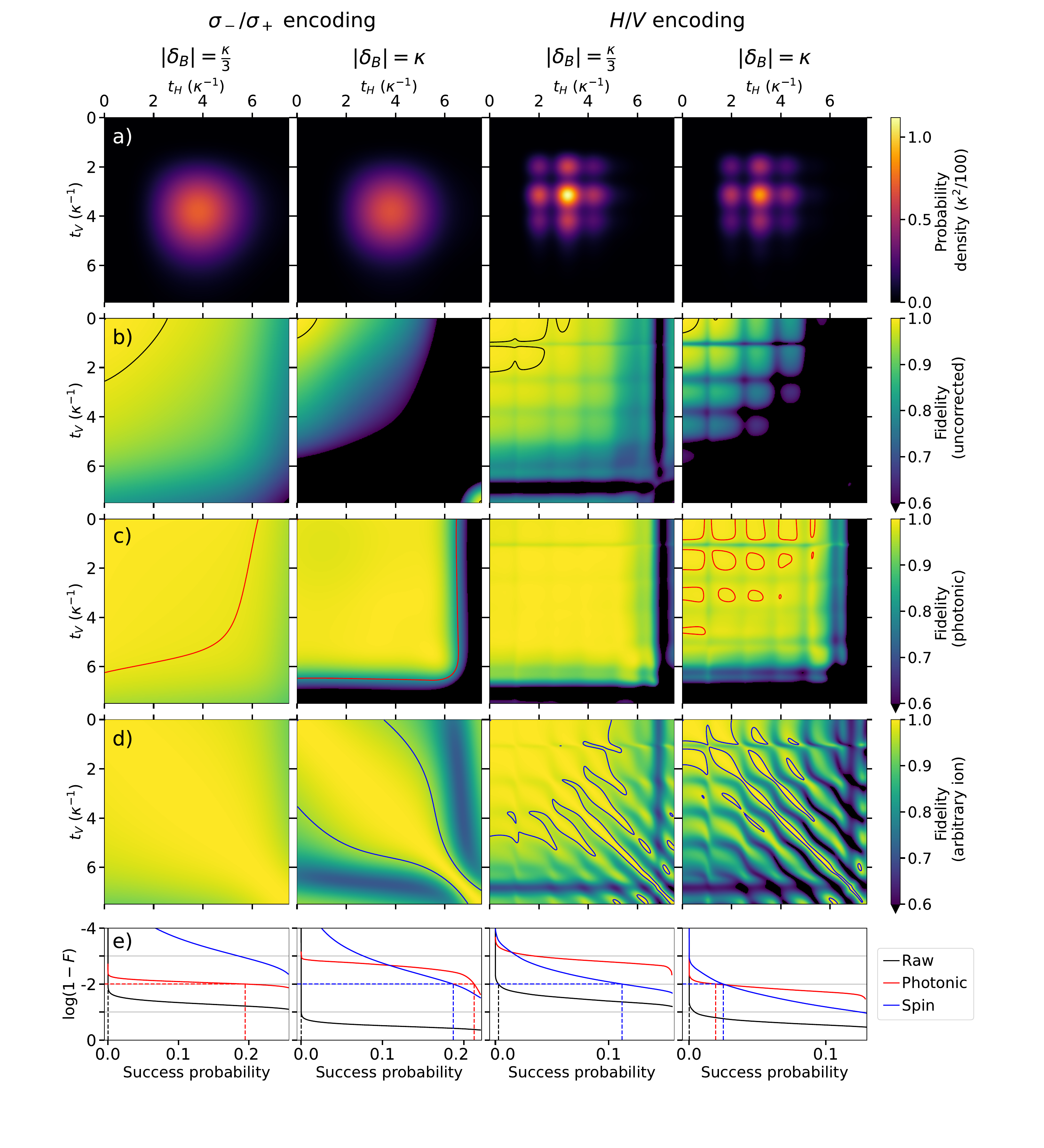}
    \caption{Comparison of fidelity performance for the correction schemes shown in Fig.~\ref{fig:hyper_plot_wavefunction_and_correction}, where the significance of the columns is equivalent. For ease of comparison, all heatmaps display data for the case of $\{(\rm{c}),(\rm{c})\}$ heralds only, but Row \textbf{(e)} summarises performance across all possible heralds. \textbf{(a)} Herald probability density versus $t_H$, $t_V$, indicating the combinations of detection times which make the most significant contribution to the entanglement success rate. Note that this probability density will be slightly altered if the photonic corrections are applied. Rows \textbf{(b)-(d)} show fidelity as a function of $t_H$ and $t_V$ for different correction schemes. The contours show the window regions that yield an average 99\% entanglement fidelity with maximum success probability; if no windowing is required to achieve this average fidelity, no contour is drawn. \textbf{(b)} Raw, uncorrected fidelity. \textbf{(c)} Fidelity after photonic corrections. \textbf{(d)} Fidelity after arbitrary post-measurement single-spin correction. Row \textbf{(e)} summarises the fidelity/rate trade-off provided by post-herald detection-time windowing applied alone or in combination with pre- or post-herald corrections. Maximum success probability is achieved when no windowing is applied, and in this case fidelity is correspondingly at a minimum; increasingly stringent selectivity of high-fidelity regions increases fidelity at the expense of success probability. The dotted lines, where visible, represent the 99\% fidelity intercept.}
    \label{fig:hyper_plot_fidelity}
\end{figure}

The results of this analysis are shown in Figs.~\ref{fig:hyper_plot_wavefunction_and_correction}~and~\ref{fig:hyper_plot_fidelity}, which will be discussed in order. Figure \ref{fig:hyper_plot_wavefunction_and_correction} shows the ion-photon wavepackets produced by the node before \textbf{(a)} and after \textbf{(b)} photonic corrections, and indicates the optimal photonic \textbf{(c)} and spin \textbf{(d, e)} corrections. The wavepackets from the birefringent nodes exhibit a time-dependent output polarisation, with the `rotated' component of the polarisation more significant for higher birefringence. In the $H/V$ scheme, the wavepacket shape is more complicated due to the bichromatic drive required. 
When photonic corrections are applied, the correlation between photon polarisation and ion spin state is almost completely restored, although the shape of the wavepacket from the birefringent node remains distorted due to the effect of birefringence on the photon generation process. 
We note that the retardance slew rates required for these photonic corrections are readily experimentally achievable for typical $\kappa$, and the three-parameter optimisation necessary to calibrate these corrections for each node may be a reasonable investment of effort for the fidelity gains this confers.

Considering instead post-herald spin-correction, the heatmaps of the optimal unitary parameters show these vary with the herald detection times in both magnitude and orientation on the Bloch sphere. In general, the correction magnitude grows with $\abs{\delta_B}$ and the variations of both the magnitude and rotation axis are more severe in the $H/V$ scheme. In comparison with the photonic corrections, these post-herald spin corrections would be much harder to implement in a practical quantum network, requiring separate calibration of the unitary for each pair of detectors and pair of nodes over all possible $(t_H,t_V)$ pairs. For the $\sigma_-/\sigma_+$ scheme, the correction angle and magnitude generally vary slowly with the two detection times, and therefore the correcting gates would not need to be calibrated over a fine grid of collection times (although we note that the gradual variation in both unitary magnitude and angle nonetheless makes optimisation challenging). For the $H/V$ case, the rapid variation in both magnitude and orientation of the correcting operation with detection times $(t_H,t_V)$ means this would require very fine-grained calibration, making the strategy all but impractical when using a bichromatic drive.

\subsubsection*{Fidelity and success probability}

Figure \ref{fig:hyper_plot_fidelity} shows how remote entanglement fidelity is impacted by birefringence before and after photonic or spin-qubit correction and herald windowing. In calculating the fidelity as a function of detection times, we assume both photons are detected on the $c$ port of the analyser, and calculate the final ion-ion output state $\psi'_{cc}$~\footnote{In general the fidelity of the entangled state can depend on the specific herald pattern, but in this case, where one output node is non-birefringent, all the $H/V$ detector combinations yield the same fidelities.}. With or without corrections applied, the fidelity is a strong function of detection time with the highest fidelities generally found at lower $(t_H, t_V)$, where the impact of birefringence remains small. The application of photonic or spin corrections enable higher fidelities to be achieved over a larger range of detection times, and are generally more beneficial for the $\sigma_-/\sigma_+$ scheme, but no scheme fully restores fidelities at all times. For the photonic corrections, the dominant limit on fidelity is not due to residual rotation of polarisation components (which are effectively corrected, as seen in Fig.~\ref{fig:hyper_plot_wavefunction_and_correction}), but to distortion in the wavepacket shape; this reduced indistinguishability leads to a post-herald ion-ion state that is not maximally entangled. For the spin qubit corrections, unit fidelity is not achieved because the post-herald ion-ion state is not in general maximally entangled due to the birefringent dynamics. The local ion spin corrections are unable to increase the degree of entanglement of the ion-ion state, however fidelity can nonetheless be improved by rotating the state to maximise overlap with the target state (see App.~\ref{app:unique and single node correction}). 

For quantum networking applications, the fidelity of remote entanglement may need to exceed a certain threshold (e.g. for error correction or efficient distillation~\cite{Deutsch:98, Nigmatullin:16}). If this fidelity cannot be achieved over the average of all detection times, one may use `windowing' (Sec.~\ref{spin_corr_win}) to trade entanglement rate for fidelity by rejecting two-photon coincidences if they would produce low fidelity entanglement. This permits high fidelities (we indicate contours of $\mathcal{F}=0.99$) to be achieved even in the presence of strong birefringence, although the entanglement rate is drastically reduced if the initial fidelities are too low. The application of active correction schemes improve these initial fidelities, increasing the acceptable window region and reducing the impact on rate. Nonetheless, if the birefringence is too high, it becomes impossible to find windowing strategies that produce entanglement with useful fidelity and rate.

\subsection{Fidelity dependence on birefringence angle}
\begin{figure}
    \centering
    \includegraphics[width=0.66\textwidth, trim=0.5cm 0.75cm 2cm 0.5cm, clip]{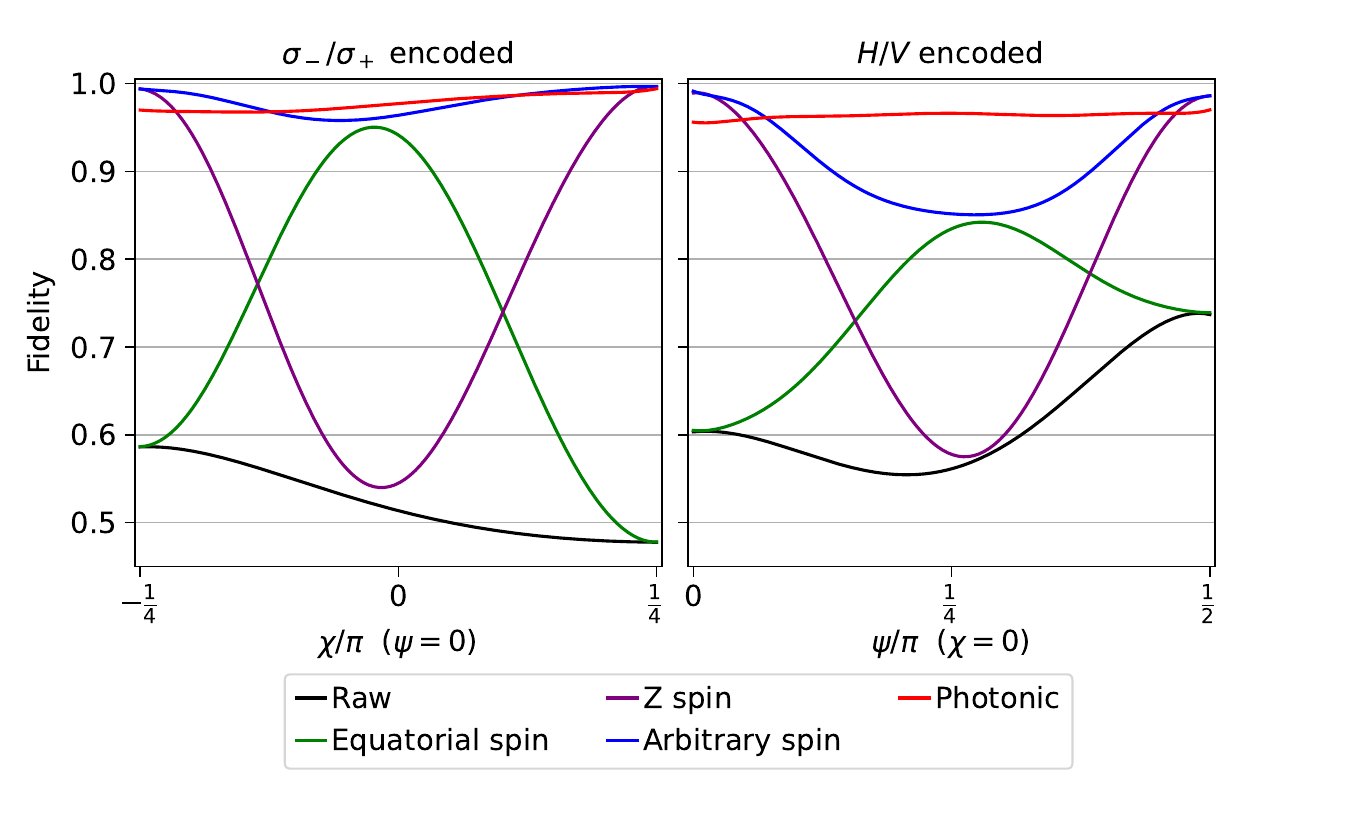}
    \caption{Dependence of fidelity on the relative orientation of the atomic decay polarisations and the birefringent cavity eigenmode polarisations for $\sigma_-/\sigma_+$ and $H/V$ schemes with a variety of correction strategies, all without detection-time windowing. The orientation of the birefringence is defined as the polarisation vector of the higher-energy polarisation eigenmode, parameterised by azimuthal ($\psi$) and altitudinal ($\chi$) angles on the Poincare sphere (see App.~\ref{app:sim_method}). In all cases, one node is ideal (zero-birefringence), while the second experiences a birefringence of magnitude $\abs{\delta_B}=\kappa$, the orientation of which is varied. As well as considering fidelity with no correction (`Raw') and after photonic and arbitrary spin corrections, we introduce two additional classes of spin correction. The `Z spin' correction stipulates that the unitary must be a rotation about the $z$ axis on the Bloch sphere, and the `Equatorial spin' that the rotation is about an (optimal) equatorial axis.  For the $\sigma_-/\sigma_+$ scheme, the fidelity with photonic and arbitrary spin corrections is generally higher and less dependent on birefringence orientation than the $H/V$ case, though it is interesting to note that the raw fidelity is generally lower. For both schemes, when the birefringent polarisation eigenmodes align with the atomic emission (the two limits of the range), the optimal arbitrary spin correction is a $z$-type unitary as the birefringence introduces a phase error with no polarisation rotation. When the birefringence is orthogonal to the atomic emission directions (centre of both plots), equatorial spin corrections become much more effective, indicating that the spin qubit populations must be partially swapped post-herald to counteract the polarisation rotation.}
    \label{fig:fidelity_dependence_on_angle}
\end{figure}

The fidelity obtained with and without correction depends not only on the magnitude of the birefringence but also its orientation with respect to the atomic emission polarisations. This is shown in Figure~\ref{fig:fidelity_dependence_on_angle}, which gives the average fidelity obtained without windowing for different correction schemes for a birefringence of magnitude $\abs{\delta_B} = \kappa$. Here, additional data is shown for cases when the ion spin unitary operation is restricted to rotations about the $z$-axis of the Bloch sphere, or to rotations about an axis in the equatorial plane of the Bloch sphere.

The first observation is that the fidelity always varies strongly with the angle of the uncontrolled birefringence (note that the range of angles considered varies between the two schemes - for more details see~\footnote{In general, birefringence of arbitrary orientation can occur in all systems, although mirror fabrication imperfections will typically introduce near-linear birefringence. In these plots we vary the orientation of the birefringence to explore the full range of projection of the birefringence on the atomic emission axes for a given birefringence magnitude, which exhibits the full range of possible fidelities to make the results complete. Because of this, the orientations of birefringence examined in the two schemes are different. In the case of the $\sigma_-/\sigma_+$ scheme, we vary the ellipticity of the birefringence, with $\psi=0$ indicating circular birefringence and $\psi=\pi/2$ linear birefringence; the tilt angle of the ellipticity is uninteresting in the case of circularly polarised atomic decay, and is thus fixed at $\chi=0$. In the case of the $H/V$ scheme, the full range of behaviour is observed when applying a linear birefringence ($\psi=0$) and varying the angle of the highest energy cavity polarisation eigenmodes from the $H$ polarised to $+$ polarised to $V$ polarised.}). Secondly, the type of post-measurement spin correction that is useful also varies with angle. When the birefringence is aligned to the atomic emission axes, the effect is purely a shift of qubit energy with no polarisation oscillations, and the correction required is thus a Pauli$-Z$ rotation. When the birefringence is aligned in a direction orthogonal on the Poincare sphere to the atomic emission axes, correction rotations about an equatorial axis on the Bloch sphere (which can counterract the impact of polarisation oscillation) are required. To correct for more general intermediate orientations of the uncontrolled birefringence, one must enact spin corrections around arbitrary (and varying) axes, making calibration challenging. Finally, it should be noted that at a birefringence angle approaching 0 or $\pi/2$, the vanishing amplitude of polarisation oscillations will make it increasingly difficult to optimise a photonic correction scheme by inspection of the output wavepacket alone, as this relies on nulling these unwanted components.

\subsection{Fidelity dependence on birefringence magnitude}
\begin{figure}
    \centering
    \includegraphics[width=0.66\textwidth, trim=0.5cm 1cm 2cm 0.5cm, clip]{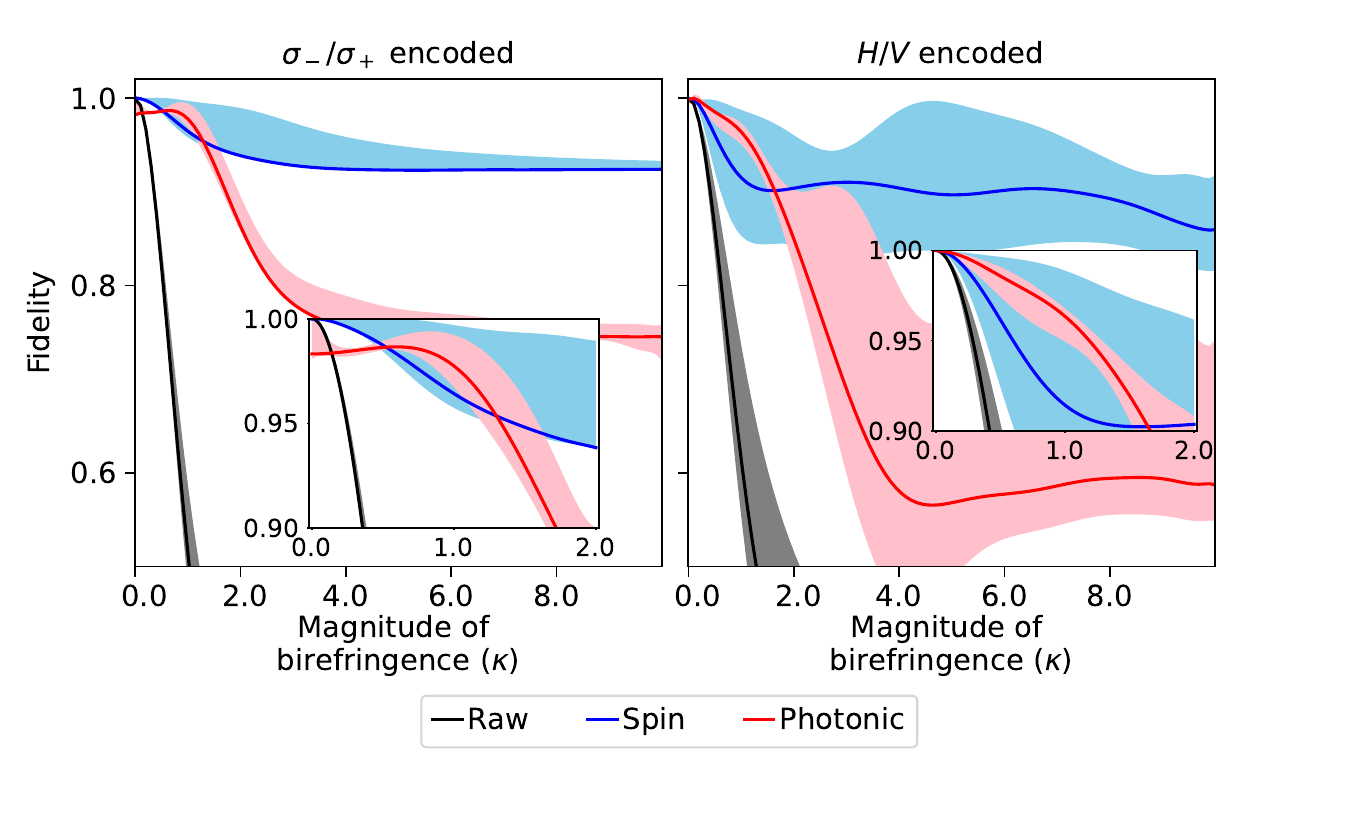}
    \caption{Fidelity versus magnitude of birefringence $\abs{\delta_B}$ for $\sigma_-/\sigma_+$ and $H/V$ schemes without detection time windowing. In each case one node is non-birefringent and the other experiences a birefringence of variable magnitude sampled from the relevant angular parameter range in Fig.~\ref{fig:fidelity_dependence_on_angle} (this covers the full range of fidelities). The fidelity for raw uncorrected heralds (black), arbitrary post-measurement spin corrections on the birefringent node (blue) and photonic corrections (red) are shown. The shaded range of these regions covers the range of fidelities over all of the angles sampled. The solid line indicates the average fidelity for all \textit{linear} birefringences, to give an estimate of the performance for the common case where linear birefringence dominates. The inset on each plot details the region of birefringence $\abs{\delta_B}<2\kappa$ where fidelities $\mathcal{F}\gtrsim0.9$, necessary for efficient entanglement distillation~\cite{Nigmatullin:16, Deutsch:98}, can potentially be maintained. For both schemes, fidelity drops off very quickly with birefringence if there are no corrections, and at very high birefringence, arbitrary spin corrections are much better than photonic corrections for maintaining reasonable fidelity. For birefringence values up to the order of the cavity decay rate, photonic corrections usually perform well, with the exception of the lowest birefringence values for the $\sigma_-/\sigma_+$ scheme. This anomaly is due to the simple Pockels cell optimisation procedure picking unhelpful orientation angles that restore that restore the spin-polarisation correlations correctly but introduce a small phase error in the final state; some mitigation of this effect would be straightforward to achieve if observed in a real system.}
    \label{fig:fidelity_dependence_on_birefringence}
\end{figure}

The previous section discussed the dependence of the entanglement fidelity on the angle of the birefringence. However, having its source in manufacturing imperfections of the cavity mirrors, the angle of the birefringence may vary greatly. It is therefore useful to consider the range of fidelities achievable for a given magnitude of birefringence, as shown in Figure \ref{fig:fidelity_dependence_on_birefringence}, providing a reasonable estimate on the maximum tolerable value of $\abs{\delta_B}$ for a given remote entanglement application. Here, once again we consider the case of remote entanglement fidelity with one ideal node and another impacted by an uncontrolled birefringence. The angle of the uncontrolled birefringence is selected from across all possible orientations on the Poincare sphere to give an indication of the dependence of the fidelity on angle.

The data highlights that without correction strategies the fidelity drops rapidly as a function of birefringence and therefore for large birefringences, aggressive windowing and a correspondingly large reduction in rate would be required to attain useful fidelities. The spin correction scheme yields superior fidelity at large birefringences, but the photonic corrections out-perform spin corrections for modest birefringence, and benefit from substantially simpler optimisation.  Both correction schemes are generally more effective for the $\sigma_-/\sigma_+$ encoding.

\subsection{Summary of results}
The central conclusion of the numerical investigation is that any significant birefringence ($\abs{\delta_B}\gtrsim0.5\kappa$) without mitigating strategies is likely to impact remote entanglement fidelity in polarisation-encoded schemes to the extent that
the fidelity is limited to below 90\%. 
The most obvious solution to these issues is to reduce the birefringence of the cavity: when $\abs{\delta_B}\ll0.1\kappa$, the associated loss of ion-ion fidelity will typically be less than $1\%$ and is unlikely to be the leading source of error in a remote entanglement generation scheme. Where this is not possible, the impact on fidelity can be reduced via correction strategies, windowing or both.

While the magnitude of birefringence remains modest ($\abs{\delta_B}<2\kappa$) it is possible to partially restore fidelity using local post-measurement operations or photonic corrections, but these come at the cost of increased experimental complexity and, in the case of spin-qubit rotations, a formidable overhead in calibrating the correction unitary operators. The effectiveness of the corrections varies, but in general corrections for the $\sigma_-/\sigma_+$ scheme are generally simpler and better-performing than those for the $H/V$ scheme. As a final method for attaining an acceptable entanglement fidelity, windowing on the detection times can strongly improve the fidelity by post-selecting only those results whose fidelity is not impacted greatly by birefringence. However, the windowing region can be complicated and the entanglement rate is often severely reduced. For very high birefringence, all strategies are likely to yield poor results.

\section{Conclusion}\label{conclusion}

Birefringent cavities are known to cause time-dependence of the polarisation of the emitted photons, but the resulting effect on the fidelity of interference-based entanglement-swapping experiments had not previously been investigated. In this paper, we have provided a detailed study of its impact on the performance of heralded, two-photon polarisation encoded schemes used to generate entanglement remotely between emitters at distant nodes. We have shown that, in the presence of birefringence, the detection of orthogonal photons no longer heralds the expected two qubit state, resulting in significantly lower entanglement fidelity. Where the birefringence approaches the cavity decay rate, the impact on fidelity is expected to be severe,
limiting the utility of such systems for quantum information applications. Many current emitter-microcavity experiments report birefringences of this magnitude~\cite{Hunger:10, Niemietz:21, Brekenfeld:20, Barrett:19, Takahashi:14, Bransdstatter:13, Uphoff:15}, so this conclusion is of some concern in designing future experiments.

When using a polarisation-encoded remote entanglement scheme, the most effective strategy is to ensure birefringence remains low enough that the fidelity impact is negligible in the context of the application. However, where higher birefringence occurs, fast electro-optic elements could be used to restore the emitted polarisations before detection, or local spin operations applied to repair the state after detection. While the photonic corrections may be simple enough to calibrate for a real experiment, we believe the calibration of the local spin corrections is likely too complicated to make this a practical solution. Aggressive windowing of the accepted herald times also yields significant improvements in fidelity, both with and without additional corrections, but this can severely impact success rate, and is not a desirable solution when remote entanglement generation is likely to be the rate-limiting step in any system deploying it.

Considering the challenges of fabricating mirrors with insignificant birefringence \cite{Garcia:18}, and the severe limitations of any of the methods with which to mitigate its impact, it is sensible to consider whether polarisation encoding of flying qubits is well-suited to large-scale quantum networks. A more effective approach may be to consider alternative photon encodings that are less sensitive to cavity birefringence, such as time-bin \cite{Luo:09} or frequency encoding \cite{Maunz:09, Connell:21}, which also benefit from compatibility with polarisation-maintaining fibre for maintaining photon indistinguishability over longer link lengths.

\section*{Acknowledgements}

The authors thank M. Keller for the conversations that inspired this study, D. Leibfried for noting an important conclusion and A. Kuhn and T. Barrett for helpful discussions. This work was funded by the United Kingdom Engineering and Physical Sciences Research Council ``Networked Quantum Information Technology'' and ``Quantum Computing and Simulation'' Hubs, and European Union Quantum Technology Flagship Project ``AQTION'' (820495). This work was supported by JST Moonshot R\&D Grant No. JPMJMS2063 and MEXT Quantum Leap Flagship Program (MEXT Q-LEAP) Grant No. JP-MXS0118067477. For the purpose of Open Access, the author has applied a CC BY public copyright licence to any Author Accepted Manuscript version arising from this submission.

\bibliographystyle{unsrt}
\bibliographystyle{apsrev4-1}

\appendix

\section{Requirements for optimal arbitrary unitary correction} \label{app:unique and single node correction}

For herald detection events at port (c), we saw the final bipartite state in the presence of birefringence took the non-maximally entangled form $\ket{\psi_{cc}'} = \ket{\uparrow\downarrow'} + \ket{\downarrow\uparrow'}$ up to a normalisation constant and with $\{\ket{\downarrow'},\ket{\uparrow'}\}$ defined in Eq. \ref{eq:primed_defs}.

The fidelity of $\ket{\psi_{cc}'}$ with the target Bell state is limited by the non-zero terms $\tilde{\beta}^{H}_V(t_V)$ and $\tilde{\beta}^{V}_H(t_H)$. We can apply local operations to nodes A and B to maximise the resulting fidelity with the target Bell state.

We first re-express $\ket{\psi_{cc}'}$ in its Schmidt-decomposed form
\begin{equation}
\ket{\psi_S} = \lambda_1 \ket{\Uparrow}_A\ket{\Downarrow}_B +\lambda_2 \ket{\Downarrow}_A\ket{\Uparrow}_B
\label{eq:starting state Schmidt}
\end{equation}
where $\{\ket{\Uparrow}_A$, $\ket{\Downarrow}_A\}$ and $\{\ket{\Uparrow}_B$, $\ket{\Downarrow}_B\}$ are orthonormal sets in the Hilbert spaces of node A and node B, respectively, and with $\lambda_{1,2} \in \mathbb{R}_+$. The target state is
\begin{equation}
\ket{\psi_T^\pm} = \frac{1}{\sqrt{2}}\left(\ket{\uparrow}_A\ket{\downarrow}_B \pm \ket{\downarrow}_A\ket{\uparrow}_B\right)
\end{equation}
with the sign determined by the detectors that click for the given herald. One might want to find the best local corrections (unitaries applied to Nodes A and B) to increase fidelity with the target state. The form of equation \ref{eq:starting state Schmidt} suggests one might wish to apply unitaries $U_A^I$ and $U_B^I$ to $\ket{\psi_S}$ by applying the following transformations

\begin{equation}
    \begin{aligned}
    \ket{\Uparrow}_A \rightarrow \ket{\uparrow}_A \, & , \, \ket{\Downarrow}_A \rightarrow \ket{\downarrow}_A\\
    \ket{\Uparrow}_B \rightarrow \pm\ket{\uparrow}_B \, & , \, \ket{\Downarrow}_B \rightarrow \ket{\downarrow}_B.
    \end{aligned}
\end{equation}

These transformations result in a state $\ket{\psi_I^\pm} = \lambda_1 \ket{\uparrow}_A\ket{\downarrow}_B \pm\lambda_2 \ket{\downarrow}_A\ket{\uparrow}_B$. Arbitrary unitaries are then applied to systems A and B (this is equivalent to applying different arbitrary transformations directly to $\ket{\psi_S}$, but the intermediate step above makes the algebra much simpler). The transformations are defined as the following

\begin{equation}
    \begin{aligned}
    \ket{\uparrow}_A &\rightarrow \cos\theta_A\ket{\uparrow}_A + \sin\theta_A e^{i\phi_A}\ket{\downarrow}_A  \, , \, \ket{\downarrow}_A \rightarrow \left(-\sin\theta_A e^{-i\phi_A}\ket{\uparrow}_A + \cos\theta_A \ket{\downarrow}_A\right)e^{i\chi_A}\\
    \ket{\uparrow}_B &\rightarrow \cos\theta_B\ket{\uparrow}_B + \sin\theta_B e^{i\phi_B}\ket{\downarrow}_B  \, , \, \ket{\downarrow}_B \rightarrow \left(-\sin\theta_B e^{-i\phi_B}\ket{\uparrow}_B + \cos\theta_B \ket{\downarrow}_B\right)e^{i\chi_B}
    \end{aligned}
\end{equation}

where each arbitrary unitary is fully specified by $\theta$, $\phi$ and $\chi$ for each node. After applying this operation, the fidelity $\mathcal{F}$ with the target state is given by

\begin{equation}
\begin{aligned}
	\mathcal{F} = \frac{1}{2}&\left(\left(\lambda_1^2+\lambda_2^2\right)\left(\cos^2\theta_A\cos^2\theta_B+\sin^2\theta_A\sin^2\theta_B - 2\Re\left(cos\theta_A\cos\theta_B\sin\theta_A\sin\theta_B e^{i\left(\phi_A-\phi_B\right)}\right)\right) \right. \\
	& \left. \pm \lambda_1\lambda_2 \left(2\Re\left(\cos^2\theta_A\cos^2\theta_B e^{i\left(\chi_A-\chi_B\right)}\right) + 2\Re\left(\sin^2\theta_A\sin^2\theta_B e^{i\left(\chi_A-\chi_B\right)} e^{-2i\left(\phi_A-\phi_B\right)}\right) \right.\right. \\
	&\left.\left. -4\Re\left(\cos\theta_A\cos\theta_B\sin\theta_A\sin\theta_B e^{i\left(\chi_A-\chi_B\right)} e^{-i\left(\phi_A-\phi_B\right)}\right)\right)\right)
\end{aligned}
\end{equation}

The maximum fidelity with $\ket{\psi_T^+}$ is obtained for $\theta_A=\theta_B$, $\phi_A=\phi_B + \pi$ and $\chi_A=\chi_B$, and with $\ket{\psi_T^-}$ for $\theta_A=\theta_B$, $\phi_A=\phi_B + \pi$ and $\chi_A=\chi_B + \pi$. This means that while there are multiple options for which unitaries to apply to the Nodes A and B to get the maximum fidelity, once the unitary applied to one qubit is specified (through its $\theta$, $\phi$ and $\chi$), the parameters for the unitary applied to the other qubit are exactly specified. In particular, one can elect to apply the arbitrary unitary on a chosen node to cancel out the initial unitary applied to that node (one of $U_A^I$ or $U_B^I$). In this case, one node has no unitary applied, and the unitary applied to the other node is completely specified. This leads to two conclusions

\begin{itemize}
    \item To obtain the maximum possible fidelity, one only needs to apply an arbitrary operation to one node (this is a property of the entangled nature of the target state)
    \item If one decides to apply a unitary to only one node, that unitary is completely specified
\end{itemize}

Finally, it is noted that the maximum fidelity $F_\text{max} = \frac{1}{2}\left(\lambda_1+\lambda_2\right)^2$ is solely a function of the Schmidt values, indicating the maximum recovered fidelity is limited by the incomplete entanglement of the two nodes after the photonic herald

\section{Simulation of a birefringent node}\label{app:sim_method}

\begin{figure}
    \centering
    \includegraphics[width=0.4\textwidth, trim=0.0cm 0.00cm 0cm 0.0cm, clip]{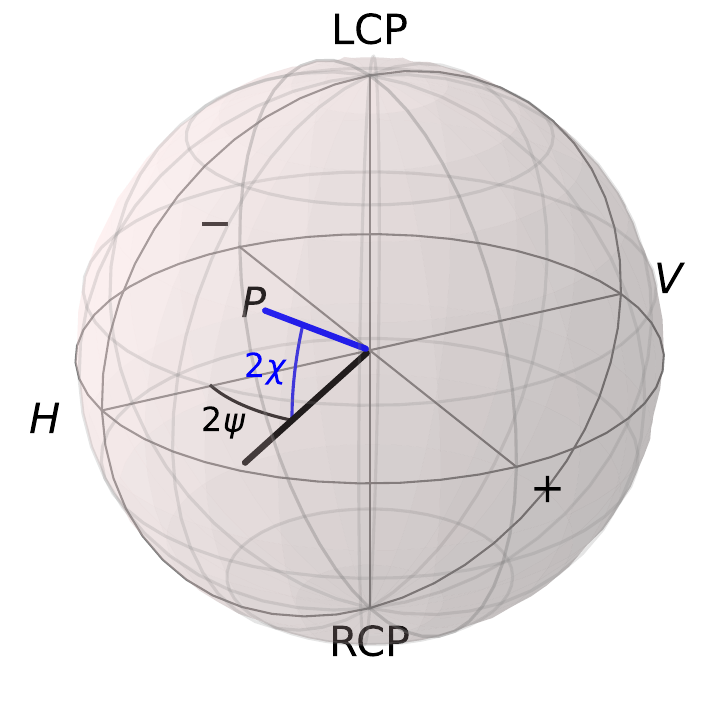}
    \caption{The Poincare sphere, depicting the angles $\psi$ and $\chi$ used in this paper. A general polarisation $P$ can be depicted as a vector from the origin to the surface of the Poincare sphere, where the angles $\psi$ and $\chi$ specify the orientation of $P$ with respect to the cardinal polarisations, where RCP and LCP are right and left circularly polarised, respectively. The propagation direction is assumed to be such that an atomic emission $\sigma_-$ corresponds to right-circularly polarised light in the $\sigma_-/\sigma_+$ scheme. Birefringence can be visualised as a vector from the origin in this coordinate system, pointing towards the highest energy polarisation eigenmode, with magnitude $\abs{\delta_B}$ that is half the frequency difference of the polarisation eigenmodes. Photons in the cavity then evolve under the Hamiltonian $H = \hbar\vec{\delta_B}\cdot\vec{\sigma}$, and evolution of the polarisation state in the cavity is analogous to evolution of a qubit state on the Bloch sphere.}
    \label{fig:poincare_sphere}
\end{figure}

Each node is modelled using four atomic states (initial state, excited state and two qubit states) and two photonic modes of the cavity ($H$ and $V$ for the $H/V$ scheme, and $\sigma_-$ and $\sigma_+$ for the $\sigma_-/\sigma_+$ scheme). The calculations use a non-Hermitian Hamiltonian approach, which assumes that all decays mechanisms cause a loss of the population from the system. In particular, it assumes that there can be no spontaneous emission that returns the atomic state to the initial state, followed by production of a cavity photon, which would cause the produced photon to have a probabilistic mixture of temporal wavepackets. This process is not considered here so that reductions in fidelity can be attributed purely to birefringence, but we note that this is an important consideration in cavity-Raman photon production schemes~\cite{Walker:20, Gao:21}.

Birefringence is introduced to the system through a Hamiltonian $H = \hbar\vec{\delta_B}\cdot\vec{\sigma}$ that acts on the polarisation subspace, where $\vec{\sigma} = \left(\sigma_1, \sigma_2, \sigma_3 \right)$ is the standard vector of Pauli matrices. $\vec{\delta_B}$ is the vector representing the birefringence on the Poincare sphere, and points from the origin in the direction of the highest energy polarisation eigenmode. The length of this vector is $\abs{\delta_B}$, meaning that the energy splitting between the two polarisation eigenmodes is 2$\hbar\abs{\delta_B}$.
To present a consistent Hamiltonian for the two schemes ($\sigma_-/\sigma_+$ and $H/V$), the mapping between the polarisation directions and the qubit states is different; this ensures the polarisation states of the emitted photons are diagonal in the Hamiltonian basis. For the $H/V$ scheme, the directions $\left(\vec{e_1},\vec{e_2},\vec{e_3}\right)$ that specify the qubit basis are $\left(\vec{e_+},\vec{e_{\text{RCP}}},\vec{e_H}\right)$ on the Poincare sphere, and for the $\sigma_-/\sigma_+$ scheme $\left(\vec{e_1},\vec{e_2},\vec{e_3}\right)=\left(\vec{e_H},\vec{e_+},\vec{e_{\text{RCP}}}\right)$, where $\vec{e_i}$ represents a unit vector in the given basis. The propagation direction is assumed to be such that $\sigma_-$ corresponds to right-circularly polarised light. In the defined basis $\vec{\delta_B}$ takes on components $\left(\delta_{B,1}, \delta_{B,2}, \delta_{B,3}\right)$,  although the angles $\psi$ and $\chi$, defining the angle of $\vec{\delta_B}$ on the Poincare sphere are unaffected by this mapping and are always defined with respect to the canonical polarisation basis as shown in Figure~\ref{fig:poincare_sphere}.

The atomic states used were $\ket{g}$, representing a stable initial atomic state and $\ket{e}$, representing an excited atomic state, as well as $\ket{\downarrow}$ and $\ket{\uparrow}$ introduced in the main text. The basis used in the $H/V$ case was, in order, $\left(\ket{g}\ket{0}, \ket{e}\ket{0}, \adag{}{{H , \text{cav}}}\ket{\uparrow}\ket{0}, \adag{}{{V, \text{cav}}}\ket{\uparrow}\ket{0}, \adag{}{{V, \text{cav}}}\ket{\downarrow}\ket{0}, \adag{}{{H, \text{cav}}}\ket{\downarrow}\ket{0}, \right)$, where $\adag{}{{x, \text{cav}}}$ is the creation operator for a photon of polarisation $x\in\{H,V\}$, with $\text{cav}$ indicating the operator acts on the photonic mode inside the cavity. These internal mode creation operators are related to the input port creation operators through $\adag{}{x} = \sqrt{2\kappa}\adag{}{{x, \text{cav}}}$. For the $\sigma_-/\sigma_+$ case, because the atomic emission polarisations are different, the basis changes to $\left(\ket{g}\ket{0}, \ket{e}\ket{0}, \adag{}{{\sigma_+ , \text{cav}}}\ket{\uparrow}\ket{0}, \adag{}{{\sigma_-, \text{cav}}}\ket{\uparrow}\ket{0}, \adag{}{{\sigma_-, \text{cav}}}\ket{\downarrow}\ket{0}, \adag{}{{\sigma_+, \text{cav}}}\ket{\downarrow}\ket{0}, \right)$. The Hamiltonian modelled was

\begin{equation}
    H = \hbar \mqty(0 & -i\Omega (t) & 0 & 0 & 0 & 0  \\ i\Omega^* (t) & \Delta_R + i\gamma & ig_\uparrow & 0 & ig_\downarrow & 0 \\ 0 & -ig_\uparrow & \frac{1}{2}\Delta_Z + \delta_{B,3} + i\kappa & \delta_{B,1}-i\delta_{B,2} & 0 & 0 \\ 0 & 0 & \delta_{B,1}+i\delta_{B,2} & \frac{1}{2}\Delta_Z -\delta_{B,3} + i\kappa & 0 & 0 \\ 0 & -ig_\downarrow & 0 & 0 & -\frac{1}{2}\Delta_Z - \delta_{B,3} + i\kappa & \delta_{B,1}+i\delta_{B,2} \\ 0 & 0 & 0 & 0 & \delta_{B,1}-i\delta_{B,2} & -\frac{1}{2}\Delta_Z + \delta_{B,3} + i\kappa)
\end{equation}
which is expressed in a frame rotating with the average energy of the cavity resonances. The initial state is chosen to be at zero detuning to fix the system onto Raman resonance. As depicted in Figure~\ref{fig:sr_yb_schemes}, $\Delta_Z$ is the frequency difference between the spin qubit states (caused by the Zeeman splitting of the manifold) and $\Delta_R=\omega_{eg}-(1/2)\left(\omega_\text{ram, H} + \omega_\text{ram, V}\right)$ is the frequency detuning of the Raman drive from the ground-excited state transition frequency. $\Omega (t)$ is the time dependent coupling of the laser drive. In the $\sigma_-/\sigma_+$ scheme, where only one tone is applied to couple symmetrically to the cavity decays, $\Omega (t)$ is linearly increasing and has a central frequency of zero in this rotating frame. In the $H/V$ scheme, the driving now contains bichromatic components, with $\Omega (t)$ = $(\Omega_{\uparrow}+\Omega_{\downarrow})$ with   $\Omega_{\uparrow,\downarrow}=\abs{\Omega_{\uparrow,\downarrow}(t)}\exp(\pm i \Delta_Z t/2)$ are the time-dependent coupling strengths of the bichromatic components of the laser drive in cavity-Raman resonance with the $\ket{\uparrow},\ket{\downarrow}$ states, and $g_\uparrow$, $g_\downarrow$ the respective cavity couplings. $\gamma = \Gamma/2$ is the decay rate of the excited state due to spontaneous emission and $\kappa$ is the cavity field amplitude decay rate. 

 Note that we do not update the drive frequencies in the presence of birefringence, and in this case both cavity eigenmodes will be symmetrically detuned from Raman resonance. Note also that where the two cavity couplings $g_\uparrow$ and $g_\downarrow$ are not equal (e.g. due to different Clebsch-Gordan coefficients), the laser drive strengths of the two frequency components are adjusted to equalise the probabilities of decay to either qubit state.

\end{document}